\documentclass{article}

\usepackage{amsmath}
\usepackage{amsfonts}
\usepackage{amssymb}
\usepackage{color}

\newcommand{\omits}[1]{}
\newcommand{\dS}{\mbox{$dS$}}
\newcommand{\AdS}{\mbox{$AdS$}}

\newcommand{\CFT}{CFT}
\newcommand{\diag}{\mathrm{diag}}

\newcommand{\id}{\mathrm{id}}
\newcommand{\vect}[1]{\mbox{\boldmath{$#1$}}}
\newcommand{\tensor}[1]{\mathbf{#1}}

\definecolor{dyellow}{rgb}{1.,0.8,.0}
\definecolor{myblue}{rgb}{.1,.1,.7}
\definecolor{dcyan}{rgb}{.0,.6,.6}
\definecolor{dmagenta}{rgb}{0.6,0.0,0.6}
\definecolor{brown}{rgb}{0.6,0.2,0.}
\definecolor{darkblue}{rgb}{.0,.0,0.5}
\definecolor{darkred}{rgb}{0.75,0.0,0.0}
\definecolor{orange}{rgb}{1.,.6,.0}
\definecolor{dorange}{rgb}{0.8,.4,.0}
\definecolor{darkgreen}{rgb}{0.0,0.6,0.0}
\definecolor{purple}{rgb}{.4,.0,.4}
\definecolor{lightgrey}{rgb}{0.7, 0.7, 0.7}

\begin{document}

\title{Conformal Triality of  de~Sitter, Minkowski and Anti-de~Sitter Spaces
\footnote{\uppercase{T}his work is supported partly by
\uppercase{NSFC} (10505004, 10375087, 90503002). }}

\author{Bin Zhou \thanks{Email: zhoub@bnu.edu.cn} \\
  \small Department of Physics, Beijing Normal University,
  \small Beijing 100875, China
\\
  Han-Ying Guo \thanks{Email: hyguo@itp.ac.cn} \\
  \small Institute of Theoretical Physics, Chinese Academy of Sciences \\
  \small P.O. Box 2735, Beijing 100080, China}

\date{December 19, 2005}

\maketitle

\begin{abstract}
We describe how conformal Minkowski, \dS- and \AdS-spaces can be united into
a single submanifold $[\mathcal{N}]$ of $\mathbb{R}P^5$. It is the set of
generators of the null cone in $\mathcal{M}^{2,4}$. Conformal transformations
on the Mink-, \dS- and \AdS-spaces are induced by $O(2,4)$ linear
transformations on $\mathcal{M}^{2,4}$. We also describe how Weyl
transformations and conformal transformations can be resulted in on
$[\mathcal{N}]$. In such a picture we give a description of how the conformal
Mink-, \dS- and \AdS-spaces as well as $[\mathcal{N}]$ are mapped from one to
another by conformal maps. This implies that a \CFT\ in one space can be
translated into a \CFT\ in another. As a consequence, the
\AdS/\CFT-correspondence should be extended.
\end{abstract}

\section{Introduction}

In this talk we show how three kinds of spaces of constant curvatures are
``unified" into a single space by conformal maps: the conformal Mink-, \dS-
and \AdS-spaces are the same nature, resulted in from a hypersurface
$[\mathcal{N}]$ of $\mathbb{R}P^5$. Here $[\mathcal{N}]$ is the quotient space
from the null ``cone" $\mathcal{N}$ of $\mathcal{M}^{2,4}$ with the vertex at
the origin. Although no metric
on $[\mathcal{N}]$ can be induced naturally from $\mathcal{M}^{2,4}$, a set of
metrics can be obtained, differing from each other by a Weyl factor. For a given
metric on $[\mathcal{N}]$, an $O(2,4)$ linear transformation on
$\mathcal{M}^{2,4}$ induces a conformal transformation.

Starting from this picture, it is not astonishing that the conformal Mink-space,
$\dS^4$ and $\AdS^4$ can be conformally mapped from one to another.
This technique can be used to translate the content of a \CFT\ from one space
to another.  Thus, if we have the \AdS/\CFT\ correspondence\cite{adscft}
between $\AdS^5$ and the conformal Mink-space, then we also have various
correspondences: $\AdS^5$ between conformal $\dS^4$, $\AdS^4$ or
$[\mathcal{N}]$.

\section{The Hypersurface $[\mathcal{N}]\subset\mathbb{R}P^5$}
\label{sect:[N]}

\subsection{The $O(2,4)$-Invariant Hypersurface of $\mathbb{R}P^5$}
\label{sect:N-[N]}

For the $(2+4)$-d Mink-space $\mathcal{M}^{2,4}$ endowed with
the inner product
\begin{equation}
  \vect{\zeta}_1\cdot\vect{\zeta}_2
  := \eta_{\hat A \hat B}\,\zeta_1^{\hat A} \zeta_2^{\hat B},
\qquad
  (\eta_{\hat A \hat B}) = \diag(1, -1, \ldots, -1, 1),
\end{equation}
where $\hat{A}, \hat{B} = 0, 1, \dots, 5$, we consider its null cone
\begin{equation}
  \mathcal{N}: \quad \vect{\zeta}\cdot\vect{\zeta} = 0,
  \qquad (\vect{\zeta} \neq 0).
\end{equation}
In $\mathcal{M}^{2,4}$ there is the standard equivalence relation $\sim$,
defined by
\begin{equation}
  \vect{\zeta}' \sim \vect{\zeta} \ \Leftrightarrow
  \ \vect{\zeta}' = c\,\vect{\zeta}
  \ \textrm{ for a nonzero } c\in\mathbb{R},
\label{def:pteql}
\end{equation}
which makes the quotient space $\mathcal{M}^{2,4}-\{0\}/\sim$ to be the
projective space $\mathbb{R}P^5$.
The equivalence class of a nonzero $\vect{\zeta}\in\mathcal{M}^{2,4}$ is denoted
by $[\vect{\zeta}]$.  Thus, $\mathcal{N}$ defines a quotient space
$\mathcal{N}/\sim\ \subset\mathbb{R}P^5$, denoted by $[\mathcal{N}]$ for
convenience.  It is obvious that $[\mathcal{N}]$ is homeomorphic to
$S^1\times S^3$.

As well known, a general linear transformation on $\mathcal{M}^{2,4}$
induces a projective transformation on $\mathbb{R}P^5$.  Since $\mathcal{N}$ is
invariant under the $O(2,4)$ linear transformations\footnote{
Strictly, $\mathcal{N}$ is invariant under the action of
$O(2,4)\times\mathbb{R}$, where $r\in\mathbb{R}$ refers to a scale product on
$\mathcal{M}^{2,4}$ by $e^r$. But the action of $\mathbb{R}$ induces the
identity transformation on $\mathbb{R}P^5$. Thus it can be safely ruled out in
our consideration.
}, 
they induce some transformations on $[\mathcal{N}]$.
In \S\ref{sect:conf-transf} we shall show how these transformations on
$[\mathcal{N}]$ can be made into a conformal transformation on $[\mathcal{N}]$.
In \S\ref{sect:Mink-dS-AdS} we shall show how these induced transformations on
$[\mathcal{N}]$ can be viewed as ``conformal transformations" on the Mink-space,
$\dS^4$ or $\AdS^4$.

Before the topic of conformal transformations is concerned, we must investigate
the problem of metric on $[\mathcal{N}]$. The metric
$ 
  \vect{\eta} = \eta_{\hat A \hat B}\ d\zeta^{\hat A}\otimes d\zeta^{\hat B}
$ 
on $\mathcal{M}^{2,4}$ cannot naturally induce a metric on $[\mathcal{N}]$.
But it is not so bad.

A curve $\gamma$ in $\mathcal{N}$ can be projected to be a curve $[\gamma]$ in
$[\mathcal{N}]$. However, the projection from $\gamma$ to $[\gamma]$ is not
one-to-one. Another curve $\gamma'$ in $\mathcal{N}$ can be projected to the
same $[\gamma]$ in $[\mathcal{N}]$ iff their parameter equations differ from
each other by a nonzero factor.  We call such two curves in $\mathcal{N}$ are
\textit{equivalent} to each other.  Given two equivalent curves
$\zeta^{\hat A} = \zeta^{\hat A}(t)$ and
$\zeta^{\hat A} = c(t)\,\zeta^{\hat A}(t)$ in $\mathcal{N}$, their line
elements, $ds^2$ and $ds'^2$, respectively, satisfy the relation
\begin{equation}
  ds'^2 = c^2\,ds^2.
\label{eq:ds'-ds}
\end{equation}

We can turn to the tangent spaces of $\mathcal{N}$ to formulate this result.
We say that two tangent vectors, $\tensor{X}\in T_{\tensor{\zeta}}\mathcal{N}$
and $\tensor{X}'\in T_{\tensor{\zeta}'}\mathcal{N}$, are \textit{equivalent} if
$[\vect{\zeta}'] = [\vect{\zeta}]$ and $\pi_*\tensor{X} = \pi_*\tensor{X}'$
where $\pi_*$ is the pull-back of the natural projection $\pi: \mathcal{N}
\rightarrow [\mathcal{N}]$.
Now suppose $\tensor{X}', \tensor{Y}'\in T_{\tensor{\zeta}'} \mathcal{N}$ are
equivalent to $\tensor{X}, \tensor{Y} \in T_{\tensor{\zeta}} \mathcal{N}$,
respectively. Then,
\begin{equation}
  \vect{\eta}(\tensor{X}', \tensor{Y}')
  = c^2\,\vect{\eta}(\tensor{X}, \tensor{Y}),
\label{eq:X'Y'-XY}
\end{equation}
where $c$ is the number in $\vect{\zeta}' = c\,\vect{\zeta}$.
This is the precise meaning that is implied, consciously or unconsciously, by
eq.~(\ref{eq:ds'-ds}).

There are two important consequences of the above result. We describe them in
\S\ref{sect:emb-Weyl} and \S\ref{sect:conf-transf}, respectively.

\subsection{Induced Metric and Weyl Transformations on $[\mathcal{N}]$}
\label{sect:emb-Weyl}

It is not only that $[\mathcal{N}]$ is a quotient manifold, but also that all
its tangent vectors can be viewed as residue classes: each residue class is a
set of equivalent tangent vectors of $\mathcal{N}$. The usual way to deal with
$[\mathcal{N}]$ is select a representative from each point in $[\mathcal{N}]$.
If all the representatives are selected perfectly, we obtain an embedding
$\phi: [\mathcal{N}] \rightarrow \mathcal{N}$ satisfying
\begin{equation}
  \pi\circ\phi = \id_{[\mathcal{N}]},
\label{eq:cnd-phi}
\end{equation}
where $\id_{[\mathcal{N}]}$ is the identity map on $[\mathcal{N}]$.  Then the
problem of selecting a representative for each tangent vector of $[\mathcal{N}]$
can be naturally solved by $\phi_*$. In this way we obtain a metric
\begin{equation}
  \tensor{g} = \phi^*\vect{\eta}
\label{def:g}
\end{equation}
on $[\mathcal{N}]$.
It is easy to see that $\tensor{g}$ is a Lorentzian metric on $[\mathcal{N}]$.

If $\phi': [\mathcal{N}] \rightarrow \mathcal{N}$ is also an embedding
satisfying $\pi\circ\phi' = \id_{[\mathcal{N}]},$ then for any $[\vect{\zeta}]
\in [\mathcal{N}]$, we have $[\vect{\zeta}] = \pi(\phi([\vect{\zeta}]))
= \pi(\phi'([\vect{\zeta}])).$ Thus there must be a nonzero real number
$\Omega([\vect{\zeta}])$ so that
\begin{equation}
  \phi'([\vect{\zeta}]) = \Omega([\vect{\zeta}])\,\phi([\vect{\zeta}]).
\label{Omega:phi'-phi}
\end{equation}
Therefore, the two embeddings $\phi$ and $\phi'$ define a nonzero function
$
  \Omega
$
on $[\mathcal{N}]$.  It is obvious that $\Omega$ is smooth.

Let $\tensor{g}' = \phi'^*\vect{\eta}$. Then it can be proved that
\begin{equation}
  \tensor{g}' = \Omega^2 \tensor{g}.
\end{equation}
That is, the consequence of the variation of embeddings is a Weyl transformation
for the induced metric on $[\mathcal{N}]$.

\subsection{Conformal Transformations on $[\mathcal{N}]$}
\label{sect:conf-transf}

In \S\ref{sect:N-[N]} we have pointed out that an $O(2,4)$ transformation $O$
on $\mathcal{M}^{2,4}$ induces a transformation $[O]$ on $[\mathcal{N}]$, well
defined by
\begin{equation}
  [O]([\vect{\zeta}]) := [O\vect{\zeta}].
\label{def:[O]}
\end{equation}
For a given $O\in O(2,4)$, $[O]$ is a diffeomorphism on $[\mathcal{N}]$.
Hence an action of $O(2,4)$ on $[\mathcal{N}]$ on the left is resulted in.
However, such an action is not effective, because it can be easily verified that
\begin{equation}
  [- E] = [E] = \id_{[\mathcal{N}]},
  \qquad \textrm{or} \qquad [-O] = [O]
\label{eq:kernel}
\end{equation}
for arbitrary $O \in O(2,4)$, where $E$ is the identity transformation on
$\mathcal{M}^{2,4}$.
It can be proved that, for an $O\in O(2,4)$, $[O] = \id_{[\mathcal{N}]}$ iff
$O = \pm\; E$. It can be also proved that the action of $O(2,4)$ on
$[\mathcal{N}]$ is transitive. So $[\mathcal{N}]$ is a homogeneous space of
$O(2,4)$.

Let $\phi: [\mathcal{N}] \rightarrow \mathcal{N}$ be an embedding as described
in \S\ref{sect:emb-Weyl}, and $O$ be an $O(2,4)$ linear transformation. For an
arbitrary $[\vect{\zeta}] \in [\mathcal{N}]$, we can set
\begin{equation}
  \vect{\zeta} := \phi([\vect{\zeta}]), \qquad
  \vect{\zeta}' := \phi([\vect{\zeta}']) = \phi([O\vect{\zeta}]),
\label{eq:zeta}
\end{equation}
which are contained in $\phi([\mathcal{N}]) \subset \mathcal{N}$ and can be
treated as representatives of $[\vect{\zeta}]$ and $[O][\vect{\zeta}]$,
respectively.  On the other hand, since
$[\vect{\zeta}'] = \pi(\vect{\zeta}') = (\pi\circ\phi)([O\vect{\zeta}])
= \id_{[\mathcal{N}]}([O\vect{\zeta}]) = [O\vect{\zeta}]$,
there must be a nonzero real number $\rho([\vect{\zeta}])$, depending on
$[\vect{\zeta}]$, such that
\begin{equation}
  \vect{\zeta}' = \rho([\vect{\zeta}])\,O\vect{\zeta}.
\label{def:rho[zeta]}
\end{equation}
In this way we obtain a nonzero function $\rho$ on $[\mathcal{N}]$.

Now let $\tensor{g}$ be the metric on $[\mathcal{N}]$ induced by the
embedding $\phi: [\mathcal{N}] \rightarrow \mathcal{N}$, as shown in
\S\ref{sect:emb-Weyl}. It can be proved that $[O]$ is a conformal
transformation:
\begin{equation}
  [O]^*\tensor{g} = \rho^2 \ \tensor{g}.
\end{equation}

So, every $O(2,4)$ linear transformation on $\mathcal{M}^{2,4}$ induces a
conformal transformation on $([\mathcal{N}], \tensor{g})$. Due to
eqs.~(\ref{eq:kernel}), the conformal group of $([\mathcal{N}], \tensor{g})$ is
the quotient group $O(2,4)/\mathbb{Z}_2$.

\section{Conformal Transformations on the Mink-Space, $\dS^4$ and
$\AdS^4$}
\label{sect:Mink-dS-AdS}

In \S\ref{sect:emb-Weyl} and \S\ref{sect:conf-transf} the representatives are
selected in a perfect way that they form a submanifold $\phi([\mathcal{N}])$
diffeomorphic to $[\mathcal{N}]$. In this section we use a not so perfect
method: only most of, but not all, points in $[\mathcal{N}]$ can find their
respective representatives, located in a hyperplane $\mathcal{P}$ of
$\mathcal{M}^{2,4}$ off $\vect{\zeta} = 0$. The resulted space
$\mathcal{P}\cap\mathcal{N}$ are Mink, $\dS^4$ or $\AdS^4$ according to
whether the normal vector $\tensor{n}$ of $\mathcal{P}$ is null, timelike or
spacelike. And ``on" these spaces there are the ``conformal transformations"
which are of great interest in physics.

\subsection{The Minkowskian Case}

When the normal vector $\tensor{n}$ is null, it can be extended to be a linear
basis $\{\tensor{e}_\mu, \tensor{n}, \tensor{l}\}$ of $\mathcal{M}^{2,4}$,
with $\tensor{e}_\mu$ for $\mu = 0, \ldots, 3$ tangent to
$\mathcal{N}$ and $\mathcal{P}$, satisfying
\begin{equation}
  \tensor{e}_\mu \cdot \tensor{e}_\nu = \eta_{\mu\nu}, \quad
  \tensor{e}_\mu\cdot\tensor{n} = 0, \quad
  \tensor{e}_\mu\cdot\tensor{l} = 0, \quad
  \tensor{l}\cdot\tensor{l} = 0, \quad
  \tensor{n}\cdot\tensor{l} = 1.
\label{eq:M-basis}
\end{equation}
It is easy to see that a point $\vect{\zeta}\in\mathcal{P}\cap\mathcal{N}$ iff
\begin{equation}
  \vect{\zeta} = x^\mu\,\tensor{e}_\mu + x^+\,\tensor{n} + R\,\tensor{l}, \qquad
  x^+ = - \eta_{\mu\nu}\,x^\mu x^\nu/(2R),
\label{eq:Mink-x}
\end{equation}
with $R$ a constant.  And it is easy to verify that $\mathcal{N}\cap\mathcal{P}$
is a Mink-space because the induced metric on it is
\begin{equation}
  ds_M^2 = \eta_{\mu\nu}\,dx^\mu\,dx^\nu.
\end{equation}

Now let us consider two equivalent curves with line elements $d\chi^2$ and
$ds_M^2$, respectively. Assume that the former is just in $\mathcal{N}$,
while the latter is in $\mathcal{P}\cap\mathcal{N}$. Then a relation similar to
(\ref{eq:ds'-ds}) can be obtained:
\begin{equation}
  d\chi^2 = (\tensor{n}\cdot\vect{\zeta}/R)^2\,ds_M^2,
\label{dchi-dsM}
\end{equation}
where $\vect{\zeta}$ is the point along the former curve.
{From} eq.~(\ref{dchi-dsM}) it can be derived that the $O(2,4)$ linear
transformations induce the so-called ``conformal transformations" on the
Mink-space\cite{twistor}.

\subsection{The $\dS^4$ and $\AdS^4$ Cases}

When the normal vector $\tensor{n}$ is timelike, the induced metric on
$\mathcal{P}$ has a signature as $\diag(1, -1, -1, -1, -1)$. Assume
$\tensor{n}\cdot\tensor{n} = 1$ and extend it to be an orthonormal basis
$\{\tensor{e}_A, \tensor{n}\,|\, A = 0, 1, \ldots, 4\}$ of $\mathcal{M}^{2,4}$.
Then $\vect{\zeta} \in \mathcal{P} \cap\mathcal{N}$ iff
\begin{equation}
  \vect{\zeta} = \xi^A\,\tensor{e}_A + R\,\tensor{n}, \qquad
  \eta_{AB}\,\xi^A \xi^B = - R^2,
\label{tilde xi}
\end{equation}
where $R$ is a positive constant and $(\eta_{AB}) = \diag(1, -1, -1, -1, -1).$
(We have carefully chosen $\tensor{n}$ in order that $R > 0$.)
Thus $\mathcal{N}\cap\mathcal{P}$ is a $\dS^4$ of radius $R$.

Let $d\chi^2$ and $ds_+^2$ be the line elements of two equivalent curves
$\gamma$ and $\gamma_+$, respectively. Again the former is just in
$\mathcal{N}$ and the latter is in $\mathcal{P}\cap\mathcal{N}$.
Then, with $\vect{\zeta}$ the point along $\gamma$, there is similarly the
relation
\begin{equation}
  d\chi^2 = (\tensor{n}\cdot\vect{\zeta}/R)^2 \, ds^2_+.
\label{dchi-ds+}
\end{equation}

Given an $O(2,4)$ linear transformation, $\gamma$ can be transformed to be
another curve $\gamma'$, lying still in $\mathcal{N}$ and equivalent to a curve
$\gamma'_+$ lying in $\mathcal{P}\cap\mathcal{N}$. Let their line elements be
$d\chi'^2$ and $ds'^2_+$, respectively. Then a similar relation to
(\ref{dchi-ds+}) holds for $d\chi'^2$ and $ds'^2_+$. The $O(2,4)$ transformation
preserves the line elements: $d\chi'^2 = d\chi^2$. Thus there will be
\begin{equation}
  ds'^2_+ = \bigg(\frac{\tensor{n}\cdot\vect{\zeta}}
    {\tensor{n}\cdot\vect{\zeta}'}\bigg)^2\,ds^2_+
\end{equation}
for $\gamma_+$ and $\gamma'_+$, where $\vect{\zeta}$ and $\vect{\zeta}'$ are
a pair of equivalent points along $\gamma$ and $\gamma'$, respectively.
Therefore, similar to the Minkowskian case, an $O(2,4)$ linear transformation
induces a ``conformal transformation" on $\dS^4$.

In general a set of Beltrami coordinates \cite{beltrami,BdS,Lu} can be assigned
to an equivalence class $[\vect{\zeta}]$.  For $\vect{\zeta}$ as in
eq.~(\ref{tilde xi}). The Beltrami coordinates for $[\vect{\zeta}]$ is
\begin{equation}
  x^\mu := R\,\xi^\mu/\xi^4, \qquad (\mu = 0, 1, 2, 3),
\end{equation}
provided that $\xi^4 \neq 0$.  In this coordinate system
\begin{equation}
  ds^2_+ = \left[\frac{\eta_{\mu\nu}}{\sigma_+(x)}
    + \frac{\eta_{\mu\alpha}\eta_{\nu\beta}x^\alpha x^\beta}
      {R^2\sigma_+(x)^2}
    \right] dx^\mu\,dx^\nu,
\quad
  \sigma_\pm(x) := 1 \mp R^{-2}\eta_{\mu\nu}x^\mu x^\nu.
\end{equation}
Here $\sigma_-(x)$ is preserved for $\AdS^4$. The Beltrami coordinates must
satisfy $\sigma_+(x) > 0$ \cite{BdS,Lu}. 
Isometries have the generic form as below\cite{BdS,Lu}:
\begin{equation}
  x'^\mu = \pm \,\frac{\sqrt{\sigma_+(a)}\,D^\mu_{\ \nu}\,(x^\nu - a^\nu)}
    {\sigma_+(a,x)},
\quad
  D^\mu_{\ \nu} := L^\mu_{\ \nu}
  + \frac{L^\mu_{\ \alpha} \,\eta_{\nu\beta}\,a^\alpha a^\beta}
    {R^2\sqrt{\sigma_+(a)}(1 + \sqrt{\sigma_+(a)})},
\label{isom}
\end{equation}
where $L = (L^\mu_{\ \nu}) \in O(1,3)$, $\pm 1 = \det L$ and the constants
$a^\mu$ satisfy $\sigma_+(a) > 0$. In the above, $\sigma_\pm(a,x) :=
1 \mp R^{-2}\eta_{\mu\nu}\, a^\mu x^\nu$, where $\sigma_-(a,x)$ is
preserved for $\AdS^4$. Other conformal transformations include
\begin{equation}
  x'^\mu = \frac{x^\mu \sqrt{1 - \beta^2}}{1 \pm \beta\sqrt{\sigma_+(x)}},
  \qquad
  (|\beta| < 1)
\end{equation}
and
\begin{eqnarray}
  x'^\mu = x^\mu - \frac{1 - \sigma_+(b,x)}{1 + \sqrt{\sigma_+(b)}} \, b^\mu
  \pm b^\mu \sqrt{\sigma_+(x)},
\end{eqnarray}
where $\pm$ corresponds to the coordinate neighborhoods where $\xi^4 > 0$ or
$\xi^4 < 0$.

When $\tensor{n}$ is spacelike, it can be similarly proved that $\mathcal{N}
\cap\mathcal{P}$ is $\AdS^4$, Similarly, $O(2,4)$ transformations induce
conformal transformations.  Beltrami coordinates can be also introduced in the
same way as on $\AdS^4$, and the conformal transformations take a similar form
as in the above.

\section{The Extension of \AdS/\CFT\ Correspondence}
\label{sect:AdS/CFT}

\subsection{The Geometric Picture of \AdS/\CFT\ Correspondence}
\label{sect:picture}

The discussion in \S\ref{sect:[N]} and \S\ref{sect:Mink-dS-AdS} reveals a
wonderful geometric picture as follows. The 4-d space $[\mathcal{N}]\cong
S^1\times S^3$ is a hypersurface of $\mathbb{R}P^5$. Although no natural metric
can be inherited from $\mathcal{M}^{2,4}$, $[\mathcal{N}]$ can be realized (by
an embedding $\phi$ as in \S\ref{sect:emb-Weyl}) as a hypersurface
$\phi([\mathcal{N}])$ of $\mathcal{N}$, enabling it to receive a metric
$\tensor{g}$ from the realization.
The variousness of realizations ends up with Weyl transformations for the
metric. Thus, $[\mathcal{N}]$ is rather a Weyl space than a spacetime, having a
vanishing Weyl tensor.
Hence theory of physics in $[\mathcal{N}]$, if exists, should be Weyl-invariant
--- at least it should be conformally invariant.

If the infinity boundary is included in the Mink-space, $\dS^4$ and $\AdS^4$,
they are also a realization of $[\mathcal{N}]$, as if the projective plane
model for $\mathbb{R}P^2$.

What soever speaking, the Mink-space, $\dS^4$ and $\AdS^4$ can be embedded into
$\mathcal{N}$, as shown in \S\ref{sect:Mink-dS-AdS}. These three kinds of
spaces, together with the perfect realizations of $[\mathcal{N}]$ as in
\S\ref{sect:[N]}, can be related to each other by the projection.
The maps from one to another are conformal maps, among which those from a
Mink/$\dS^4$/$\AdS^4$ to a Mink/$\dS^4$/$\AdS^4$ are of special
interest, which will be discussed elsewhere\cite{gz05}.%

Using the above conformal maps, a \CFT\ on the Mink-space can be transferred
to be \CFT s on both $\dS^4$ and $\AdS^4$, and vice versa. This fact can be
summarized as the conformal triality of Mink-, \dS- and \AdS-spaces. In fact,
a \CFT\ on any of these spaces is a \CFT\ on $([\mathcal{N}], \tensor{g})$.

Topologically $\AdS^5$ can be viewed as an open region in $\mathbb{R}P^5$,
consisting of timelike 1-d linear subspaces of $\mathcal{M}^{2,4}$. In this
sense $[\mathcal{N}] = \partial(\AdS^5)$. If the \AdS/\CFT\ %
correspondence\cite{adscft} is correct, then we can say that the corresponding
\CFT\ is on the Mink-space, on $\dS^4$, on $\AdS^4$, on
$([\mathcal{N}], \tensor{g})$. Thus we might have as many \AdS/\CFT\ %
correspondences as possible.

The \AdS/\CFT\ correspondence for higher dimensions can be also conjectured in
the similar geometric picture.

\subsection{Null Geodesics}

As we know, up to re-parameterizations, null geodesics are invariant
under conformal transformations and Weyl transformations. The null geodesics
can also be illustrated in a geometric picture.

Suppose $[\vect{\zeta}_0]$ and $[\vect{\zeta}_1]$ are two distinct points in
$[\mathcal{N}]$.  Then $\vect{\zeta}_0$ and $\vect{\zeta}_1$ are two linearly
independent null vectors, spanning a 2-d linear subspace (plane) $\Sigma$ in
$\mathcal{M}^{2,4}$. If, in addition,
\begin{equation}
  \vect{\zeta}_0 \cdot \vect{\zeta}_1 = 0,
\end{equation}
then the whole $\Sigma$ except the origin $0$ is contained in $\mathcal{N}$.
Thus $\Sigma\cap\mathcal{P}\subset\mathcal{N} \cap\mathcal{P}$, no matter
whether the latter is the Mink-, \dS- or \AdS-space. Obviously,
$\Sigma\cap\mathcal{P}$ is a null straight line. If, in addition, we assume that
$\vect{\zeta}_0$ and $\vect{\zeta}_1\in \mathcal{P}$, then the equation of
$\Sigma\cap\mathcal{P}$ reads
\begin{equation}
  \vect{\zeta}(\lambda) = (1-\lambda)\,\vect{\zeta}_0 + \lambda\,\vect{\zeta}_1.
\label{eq:null-strl}
\end{equation}

For the 2-d linear subspace $\Sigma\subset\mathcal{M}^{2,4}$, an antisymmetric
tensor
\begin{equation}
  \vect{\omega} := \vect{\zeta}_0\otimes\vect{\zeta}_1
  - \vect{\zeta}_0\otimes\vect{\zeta}_1
\end{equation}
can be defined in terms of its linear basis $\{\vect{\zeta}_0,
\vect{\zeta}_1\}$. If the linear basis of $\Sigma$ is changed, then the
antisymmetric tensor $\vect{\omega}'$ corresponding to the new basis is
proportional to $\vect{\omega}$. In fact, $\vect{\omega}$ can be treated
to be something like the area 2-form of $\Sigma$.

Meanwhile, for the straight line (\ref{eq:null-strl}), a 6-d angular
momentum tensor
\begin{equation}
  \vect{\mathcal{L}} := \vect{\zeta}\otimes\frac{d\vect{\zeta}}{d\lambda}
  - \vect{\zeta}\otimes\frac{d\vect{\zeta}}{d\lambda}
\end{equation}
can be defined. Substituting eq.~(\ref{eq:null-strl}) into the above, we find
that the angular momentum is conserved:
\begin{equation}
  \vect{\mathcal{L}} = \vect{\omega}.
\end{equation}

It is very intuitive and can be proved that, in the Mink, \dS\ and
\AdS\ cases, $\Sigma\cap\mathcal{P}$ is a null geodesic. The 6-d angular
momentum can be expressed in terms of the 4-d angular momentum and the
4-momentum of the massless particle.

In order to see what it looks like in the Minkowski or Beltrami coordinates, we
consider two special cases. In the first case, $\mathcal{P}$ is a null
hyperplane $\zeta^- = R$, where
$\zeta^\pm = \frac{1}{\sqrt{2}}\, (\pm\zeta^4 + \zeta^5)$.
Substitution of eq.~(\ref{eq:Mink-x}) to (\ref{eq:null-strl}) yields
\begin{equation}
  x^\mu = (1 - \frac{\tau}{R})\,x_0^\mu + \frac{\tau}{R}\,x_1^\mu,
\qquad
  \tau = R\,\frac{\lambda\,\zeta_1^-}
    {(1 - \lambda)\,\zeta_0^- + \lambda\,\zeta_1^-}.
\label{eq:x-tau}
\end{equation}
Let's write the energy-momentum and angular momentum as
\begin{equation}
  P^\mu = m\, \frac{dx^\mu}{d\tau}, \quad
  L^{\mu\nu} = x^\mu P^\nu - x^\nu P^\mu,
\label{def:PL}
\end{equation}
respectively and formally introduce
\begin{equation}
  P^+ = m\,\frac{dx^+}{d\tau}, \quad
  L^{\mu +} = - L^{+\mu} = x^\mu\,P^+ - x^+\,P^\mu,
\end{equation}
with $x^+$ as shown in eqs.~(\ref{eq:Mink-x}).  Then it can be verified that
\begin{equation}
  \mathcal{L}^{\mu\nu} = \frac{R}{m}\,L^{\mu\nu},
\quad%
  \mathcal{L}^{\mu +} = \frac{R}{m}\,L^{\mu +},
\quad%
  \mathcal{L}^{\mu -} = - \frac{R^2}{m}\,P^\mu,
\quad%
  \mathcal{L}^{+-} = - \frac{R^2}{m}\,P^+.
\end{equation}
In the second case, we take $\mathcal{P}$ as $\zeta^5 = R$. In the
corresponding Beltrami coordinate system, the equation of
$\Sigma\cap\mathcal{P}$ is still of the form (\ref{eq:x-tau}),
only that $\tau$ is no longer an affine parameter. However, with
the momentum and angular momentum still defined as in
(\ref{def:PL}), they are conserved quantities, and there will be
\begin{eqnarray}
  & & \mathcal{L}^{\mu\nu} = \frac{\zeta_0^4 \zeta_1^4}{mR}\,L^{\mu\nu},
  \quad \mathcal{L}^{\mu 4} = - \frac{\zeta_0^4 \zeta_1^4}{m}\,P^\mu,
\\
  & & \mathcal{L}^{\mu 5} = \mp \frac{\zeta_0^4 \zeta_1^4}{m}\,\eta^{\mu\nu}\,
    \sigma(x)^{\frac{3}{2}}\,g_{\nu\rho}(x)\,P^\rho,
\\
  & & \mathcal{L}^{45} = \mp \frac{\eta_{\mu\nu}\,x^\mu}{\sqrt{\sigma(x)}}\,
    \frac{\zeta_0^4 \zeta_1^4}{mR}\,P^\nu.
\end{eqnarray}
In the above, $\mp$ is opposite to the sign of $\xi^4$. If the normal vector of
$\mathcal{P}$ is spacelike, the results are similar to the above.

\section{Conclusion}

{From} the null cone $\mathcal{N}\subset\mathcal{M}^{2,4}$, we can construct
the Mink-space, $\dS^4$ and $\AdS^4$ on which the induced action of $O(2,4)$ is
conformal. When $\mathcal{M}^{2,4}$ is viewed as the homogeneous space of
$\mathbb{R}P^5$, $[\mathcal{N}] := \mathcal{N}/\sim$ is the conformal
(extension of the) Mink-, \dS- or \AdS-space. Various metrics can be endowed
on $[\mathcal{N}]$, differing from one another by Weyl transformations. For a
given metric among them, the $O(2,4)$
transformations induce conformal transformations on $[\mathcal{N}]$. Since
$[\mathcal{N}]$, the Mink-, \dS- and \AdS-spaces are related by conformal maps,
a \CFT\ in one space results in a \CFT\ in each of other spaces. Therefore the
\AdS/\CFT\ correspondence could be extended to all these spaces. The same idea
can be generalized to higher dimensions.

We have shown some evidence that the role of Beltrami coordinates on
\dS/\AdS-spaces is similar to that of the Mink- coordinates. In fact,
in the study of kinematics and dynamics on \dS/\AdS-spaces\cite{BdS,Lu},
it is also revealed.  The similarity is so strong that special relativity can
be appealed for on \dS/\AdS-spaces\cite{BdS,Lu}.

\section*{Acknowledgments}

Part of the contents of this talk is based on the co-operation\cite{gz05} of Profs.~Z.~Xu,
C.~G.~Huang and Dr.~Y.~Tian. Many thanks are expressed to them. We are also
pleasant to thank Profs.~Q.~K.~Lu, S.~K.~Wang, K.~Wu and X.~C.~Song for their
helpful discussions. B.~Zhou would like to express his special thanks to
Nankai Institute of Mathematics for the accommodation, especially to
Profs.~C.~Bai and W.~Zhang.

\end{document}